\documentclass[aps,prl,twocolumn,superscriptaddress,showpacs]{revtex4-1}

\usepackage{graphicx}
\usepackage{amsmath}
\usepackage[usenames,dvipsnames]{color}
 \usepackage{pdfcolmk}

\begin{document}

\title{Photonic Counterparts of Cooper Pairs}

\author{Andr\'e Saraiva}
\thanks{These authors contributed equally to the present work.}
\affiliation{Instituto de F\'\i sica, UFRJ, CP 68528, Rio de Janeiro, RJ, 21941-972, Brazil.}
\author{Filomeno S. de Aguiar J\'unior}
\thanks{These authors contributed equally to the present work.}
\affiliation{Departamento de F\'\i sica, ICEx, Universidade Federal de Minas Gerais. Av. Antonio Carlos, 6627, Belo Horizonte, MG, 31270-901, Brazil.}
\author{Reinaldo de Melo e Souza}
\affiliation{Instituto de F\'\i sica, UFRJ, CP 68528, Rio de Janeiro, RJ, 21941-972, Brazil.}
\author{Arthur Patroc\'\i nio Pena}
\affiliation{Departamento de F\'\i sica, ICEx, Universidade Federal de Minas Gerais. Av. Antonio Carlos, 6627, Belo Horizonte, MG, 31270-901, Brazil.}
\author{Carlos H. Monken}
\affiliation{Departamento de F\'\i sica, ICEx, Universidade Federal de Minas Gerais. Av. Antonio Carlos, 6627, Belo Horizonte, MG, 31270-901, Brazil.}
\author{Marcelo F. Santos}
\affiliation{Instituto de F\'\i sica, UFRJ, CP 68528, Rio de Janeiro, RJ, 21941-972, Brazil.}
\author{Belita Koiller}
\affiliation{Instituto de F\'\i sica, UFRJ, CP 68528, Rio de Janeiro, RJ, 21941-972, Brazil.}
\author{Ado Jorio}
\affiliation{Departamento de F\'\i sica, ICEx, Universidade Federal de Minas Gerais. Av. Antonio Carlos, 6627, Belo Horizonte, MG, 31270-901, Brazil.}

\date{\today}
		
\begin{abstract}
The microscopic theory of superconductivity raised the disruptive idea that electrons couple through the elusive exchange of virtual phonons, overcoming the strong Coulomb repulsion to form Cooper pairs. Light is also known to interact with atomic vibrations, as for example in the Raman effect.
We show that photon pairs exchange virtual vibrations in transparent media, leading to an effective photon-photon interaction identical to that for electrons in BCS theory of superconductivity, in spite of the fact that photons are bosons. In this scenario, photons may exchange
energy without matching a quantum of vibration of the medium. As a result, pair correlations for photons scattered away from the Raman resonances are expected to be enhanced. Experimental demonstration of this effect is provided here by time correlated Raman measurements in different media. The experimental data confirm our theoretical interpretation of a photonic Cooper pairing, without the need for any fitting parameters.
\end{abstract}

\maketitle

The Raman effect is explored across disciplines to expose the chemical and structural characteristics of molecules and solids. Conversely, this interaction of light and matter is utilized for the generation of non-linear effects on photons, a key resource for emerging quantum technologies. The interaction of light with the motional excitations of matter is a resource for quantum applications under intensive recent investigation~\cite{Kuzmich,Caspar,Chen,Reim,duan,Lee,Lee2,Ding2,Roelli,riedinger}.

\begin{figure}
\includegraphics[width=\columnwidth]{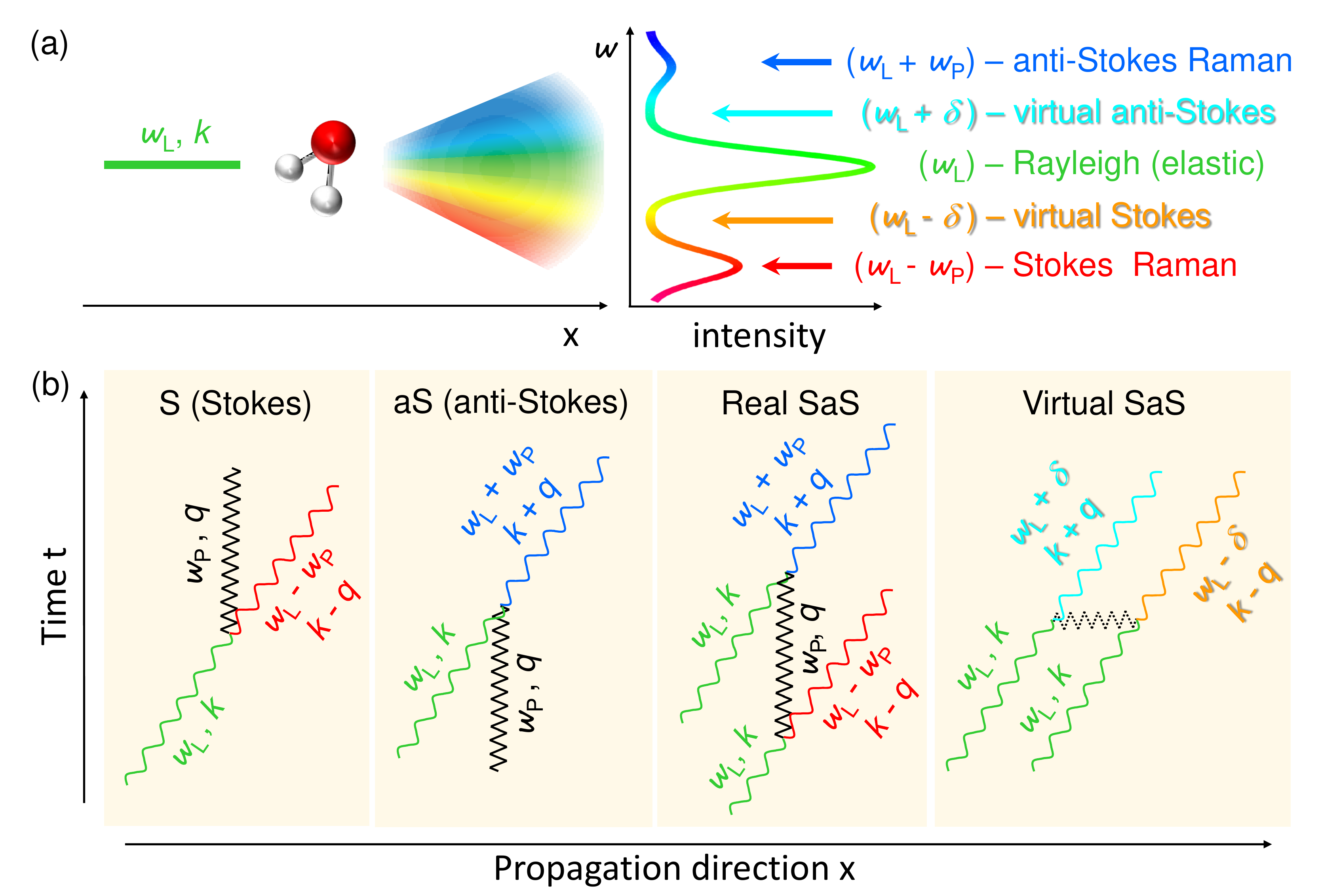}
\caption{(a) Photons from a monochromatic laser are scattered by a molecule into different angles and energies. Lateral features are found in the spectrum, which correspond to Stokes (redshifted) and anti-Stokes (blueshifted) Raman peaks. (b) Schematically represented Raman scattering processes through phonon mediated photon-photon interactions, including the usual Stokes (S) and anti-Stokes (aS) scattering, and a two-step correlated real process (Real SaS) or an exchange of virtual phonons (Virtual SaS).}
\label{fig1}
\end{figure}

The Raman effect~\cite{raman} is described in terms of a shift in the photon energy due to the inelastic scattering with matter [Fig.\,\ref{fig1}(a)]. In the Stokes process [S diagram in Fig.\,\ref{fig1}(b)], a redshifted photon emerges as part of its energy is converted into a quantum of atomic vibration (phonons for solids; vibrons for molecules or molecular clusters). A blueshift of the photons by an anti-Stokes process [aS diagram in Fig.\,\ref{fig1}(b)] is less frequent, since it requires the availability of a vibrational excitation.

A photon may also absorb the vibrational excitation generated by a previous Stokes scattering. While such a combined Stokes--anti-Stokes process [Real SaS diagram in Fig.\,\ref{fig1}(b)] is often obfuscated by the usual single scattering mechanisms, this second-order effect dominates the anti-Stokes production in the absence of native thermal quanta of vibration~\cite{klyshko77,parra}. This can be observed in crystalline systems~\cite{Lee} and molecules~\cite{kasper2} with high vibrational energies (strongly bound atoms of low mass). The yield of the SaS process is quantified by the cross-correlation between the number of Stokes photons $n_S$ and anti-Stokes photons $n_{aS}$ detected after some time delay $\tau$, defined as $g^{(2)}_{S,aS}(\tau)=\langle n_S(0) n_{aS}(\tau)\rangle/(\langle n_S(0)\rangle \langle n_{aS}(\tau)\rangle$). Recent measurements in different materials~\cite{Lee2,riedinger,kasper2} indicated a time correlation at zero delay $g^{(2)}_{S,aS}(\tau=0)$ beyond the maximum classical correlation, indicating that this is a phenomenon governed by quantum mechanics~\cite{walls}. This suggests that virtual processes [Virtual SaS diagram in Fig.\,\ref{fig1}(b)] may occur~\cite{4wave}.

In this letter, we show that these virtual SaS processes are responsible for the formation of time correlated Stokes-anti-Stokes photon pairs, which are photonic counterparts of Cooper pairs. We demonstrate this effect theoretically and experimentally for eight different materials at room temperature. Exceedingly high time correlations are obtained regardless of the molecule under study, as long as the photons are coupled by the exchange of virtual vibrations. The theoretical correlation estimated from the BCS Hamiltonian agrees with experimental data of time correlations resolved in energy throughout the whole vibrational spectrum without any fitting parameters.

The exchange of virtual phonons is the mechanism for Cooper pair formation in the BCS theory of superconductivity~\cite{BCS}. The BCS hamiltonian rederived for the case of bosons~\cite{SM} is written as
\begin{equation}
H_{int}^{pp}=\sum_{\mathbf{k},\mathbf{k}',\mathbf{q}}\frac{M_{\mathbf{q}}^2\nu_{\mathbf{q}}}{\hbar[(\omega_{\mathbf{k}}-\omega_{\mathbf{k}-\mathbf{q}})^2-\nu_{\mathbf{q}}^2]}
b^{\dagger}_{\mathbf{k}-\mathbf{q}}b^{\dagger}_{\mathbf{k}'+\mathbf{q}} \,
b_{\mathbf{k}}b_{\mathbf{k}'}
\label{hpp}
\end{equation}
where $\nu_{\mathbf{q}}$ is the frequency of the vibrational mode exchanged among the photons and $\omega_{\mathbf{k}}$ is the dispersion relation of the free photon. The momenta ${\mathbf q}$ relate to the momentum transferred to the center of mass in the case of molecules or to the crystalline momentum of the phonon for crystals. The coupling strength $M_{\mathbf{q}}$ is independent of the photon mode ${\mathbf k}$, assuming constant values for each vibrational mode ${\mathbf q}$. Therefore, the process described in Eq.~(\ref{hpp}) consists of two \emph{quanta} of the incident laser pulse with wavevectors $\mathbf{k}$ and $\mathbf{k}'$, that are inelastically scattered following momentum conservation. The initial laser field is in the coherent state $|\alpha_L\rangle$. Hence, the rate of Raman photons generated can be evaluated from the simpler effective Hamiltonian~\cite{SM} $H_{int}=\sum_{\mathbf{k},\mathbf{q}}\Delta(\mathbf{k},\mathbf{q})b^{\dagger}_{\mathbf{k}+\mathbf{q}}b^{\dagger}_{\mathbf{k}-\mathbf{q}}$ with
\begin{equation}
\Delta(\mathbf{k},\mathbf{q})=\frac{M_{\mathbf{q}}^2\alpha_L^2\nu_{\mathbf{q}}}{\hbar[(\omega_{\mathbf{k}}-\omega_{\mathbf{k}-\mathbf{q}})^2-\nu_{\mathbf{q}}^2]} \, . \label{gap}
\end{equation}
Equation~\ref{gap} is the analogue of the superconducting gap in the mean field approximation of the BCS hamiltonian
-- as long as the laser field is coherent, this transformation is exact. When negative, the interaction between the photons is attractive and may be tuned by the intensity of the laser, which sets the value of $\alpha_L^2$. From Eq.~(\ref{gap}) we see that attraction happens when the virtual vibration has less energy than the real process demands, and repulsion occurs for the opposite case. Either cases may be observed within the same material, selecting with filters which pairs of photons will be measured.

As long as the two photons scatter at the molecules within a certain small time interval $\Delta t$, the energy exchanged by the photons may differ from the quantum of vibration $h\nu_q$ by as much as $\Delta E\approx \hbar/\Delta t$. Experimental evidence of Cooper pairing is investigated here by measuring simultaneous pairs of photons scattered at shifts that violate energy conservation at each separate step, while conserving total energy in the complete process.

\begin{figure*}
\includegraphics[trim={0 2cm 0 3cm},clip,width=\textwidth]{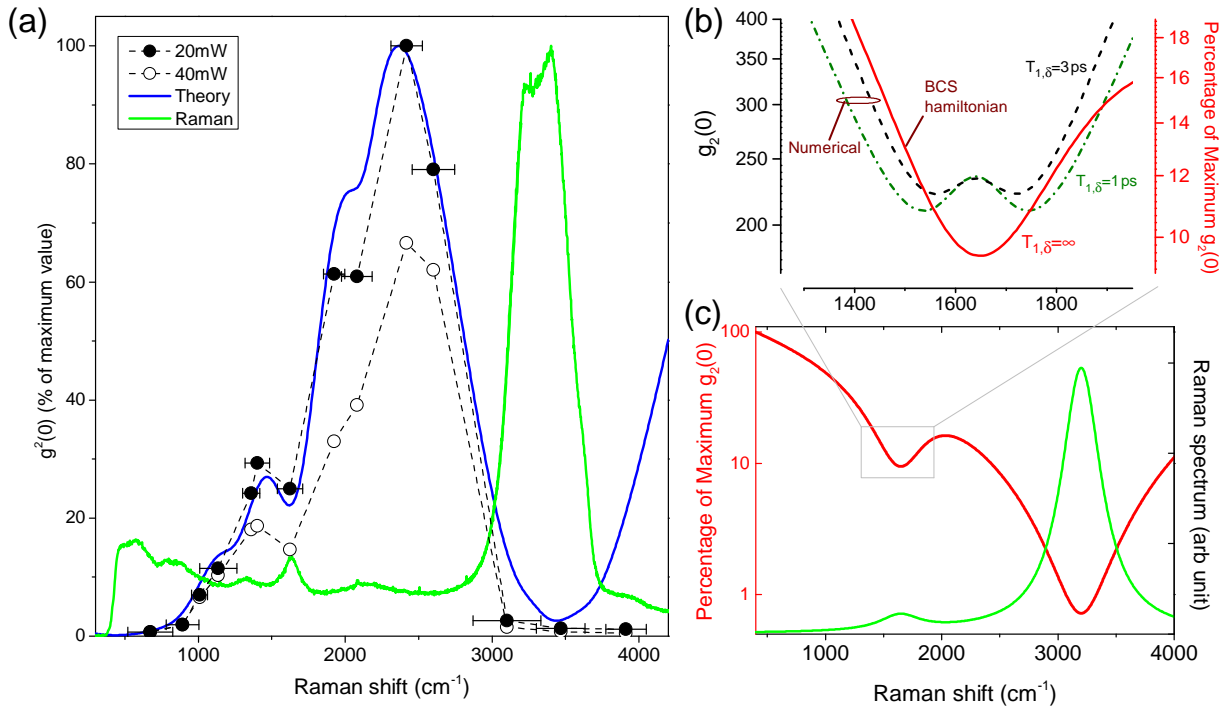}
\caption{Stokes-anti-Stokes correlation function $g^{(2)}_{S,aS}(0)$ for different values of Raman shifts. (a) Experimental values are shown as symbols connected by dotted-lines, at two excitation laser powers (see legend). Each data point was obtained with a pair of bandpass filters equally spaced from the excitation laser line towards the Stokes and anti-Stokes shifts~\cite{SM}. The green solid line represents the Raman spectrum of water, as obtained with the experimental apparatus~\cite{SM}. The absolute values for correlation are highly sensitive to instrumental performance, thus we provide here the percentage $g^{(2)}_{S,aS}(0)$ with respected to the highest observed value, all obtained with the same instrumental apparatus (except for the Stokes and anti-Stokes bandpass filters). The values were also corrected to account for the dependence of the measured $g^{(2)}_{S,aS}(0)$ with the band-overlap between Stokes and anti-Stokes filters, and the bandpass overlaps are shown as horizontal bars, indicated only in one set of data. For the as measured $g^{(2)}_{S,aS}(0)$ values, see~\cite{SM}. The solid blue line is the correlation calculated within a perturbative BCS approach [Eq.~(\ref{timeevolution})], using as input the experimental Raman spectrum, without fitting parameters. (b, c) The correlation function $g^{(2)}_{S,aS}(0)$ is calculated (solid red line) within a perturbative BCS approach for a simplified model with only two vibrational modes of the water molecule [green line in (c)], shown near the real scattering condition $(\omega_{\mathbf{k}}-\omega_{\mathbf{k}-\mathbf{q}})\approx$ 1640 cm$^{-1}$. We also present as dashed and dash-dotted lines the numerical solutions to the problem of a single vibrational mode, including two values for the relaxation time $T_1$.}
\label{fig2}
\end{figure*}

The symbols connected by dotted lines in Fig.\,\ref{fig2}(a) show the normalized $g^{(2)}_{S,aS}(0)$ measured from distilled water, for values of Raman shifts including both spectrum maxima (generating pairs through the exchange of both real and virtual vibrations) and away from these maxima (which are generated solely by virtual vibrations). The absolute measured $g^{(2)}_{S,aS}(0)$ values can be found in~\cite{SM}, but the focus here is on the relative values among the data. Filled and open circles represent measurements performed at two different excitation laser powers, namely $\sim 20$\,mW (filled circles) and $\sim 40$\,mW (open circles) at the sample. The $g^{(2)}_{S,aS}(0)$ is generally 40\% lower for the higher excitation laser fluency, as expected~\cite{parra}. Superimposed to the $g^{(2)}_{S,aS}(0)$ data points is the corresponding Stokes Raman spectrum from water (solid green line), allowing the identification of $g^{(2)}_{S,aS}(0)$ measurements in- and off-resonance with vibrational states. Here we observe an enhancement of the non-classical $g^{(2)}_{S,aS}(0)$ values for filters that do not overlap with a real vibrational mode frequency, thus providing experimental proof of the virtual SaS process [Fig.\,\ref{fig1}(b)].

For classical sources of light, the cross-correlation $g^{(2)}_{S,aS}(0)$ is bound by the Cauchy-Schwarz inequality $[g^{(2)}_{S,aS}(0)]^2 \leq g^{(2)}_{S,S}(0) \times g^{(2)}_{aS,aS}(0)$, where $g^{(2)}_{S,S}(0)$ is the autocorrelation of the Stokes signal, and similarly for $g^{(2)}_{aS,aS}(0)$.  This inequality is strongly violated (by orders of magnitude) for both real (see~\cite{kasper2}) and virtual (see~\cite{SM}) SaS processes.  

When the band of the filters is shifted away from the excitation laser energy, the virtual SaS process dominates, as evidenced by the observed enhancement of $g^{(2)}_{S,aS}(0)$ values in Fig.\,\ref{fig2}(a). This increase is observed only for symmetrically shifted filters -- an asymmetric shift suppresses the correlation. The $g^{(2)}_{S,aS}(0)$ decays when the bandpass filters match the water bending vibrational mode at 1640\,cm$^{-1}$, i.e. when the real SaS process takes place. Another mild decay in $g^{(2)}_{S,aS}(0)$ is observed when measuring the SaS process in resonance with the combination mode at 2110\,cm$^{-1}$ (bending plus intermolecular vibration~\cite{Vallee03}), which is a rather weak Raman scatterer. A precipitous $g^{(2)}_{S,aS}(0)$ drop is observed when measuring the SaS process in resonance with the high intensity 3000-3500\,cm$^{-1}$ Raman band. Therefore, the higher the differential Raman cross-section, the higher the contribution of the processes discussed in Figs.~\ref{fig1}(b), which actually washes out the $g^{(2)}_{S,aS}(0)$ values. Finally, above 3500\,cm$^{-1}$ the SaS process is no longer observed and $g^{(2)}_{S,aS}(0) \rightarrow 2$, which is the value for thermal light. Above 4,000\,cm$^{-1}$ our measurements for $g^{(2)}_{S,aS}(0)$ are no longer conclusive due to losses in the optical devices and photon counters.

A quantitative comparison between the predictions of hamiltonian in Eq.~(\ref{hpp}) and the experimental data may be obtained by calculating the final state of the photonic field using a first order perturbative approximation to the time evolution operator
\begin{equation}
U(t) \cong {\mathbf 1}  + i {\frac{ H_{int} \,\, \delta t}{\hbar}} ~. \label{timeevolution}
\end{equation}
This evolution takes the initial coherent state of the laser $|\alpha_L\rangle$ into a largely unaffected outgoing laser field superimposed to a small amplitude of states that contain a pair of photons symmetrically shifted from the laser frequency $\omega_L$. Singling out the states corresponding to photon pairs, we obtain the correlation functions. The relative intensity of the scattering amplitudes for the various vibrational and librational modes was estimated from the Stokes Raman spectrum [green line in Fig.~\ref{fig2}(a)]. This spectrum may also be used for a more accurate determination of the rate of uncorrelated coincidence events $\langle n_S\rangle\times\langle n_{aS}\rangle$, in which the anti-Stokes intensity is estimated from the Stokes intensity and the Bose-Einstein distribution of vibrational excitations (a better description may be obtained following Ref.~\cite{parra}).

The resulting estimate of $g^{(2)}_{S,aS}(0)$ is shown as a solid blue line in Fig.~\ref{fig2}(a) for the case of a 20mW excitation laser power (equivalent agreement is obtained for the 40 mW data). Correlations peak around 2400 cm$^{-1}$ because this range of shifts is the furthest from any vibrational modes of water. The only entry in theory is the Raman spectrum (green line), and the excellent agreement between theory (blue line) and experiment (circles connected by dashed-lines) without recurring to fitting parameters proves that virtual phonons are indeed responsible for the formation of the correlated SaS photons.

The intricate behaviour of the correlation due to virtual processes stems from the relatively low symmetry of the water molecule, leading to a variety of Raman features. For a single vibrational mode, the Hilbert space of phonons and photons is small enough to allow a non-perturbative numerical solution of the density matrix time evolution~\cite{SM}. In this restricted model we may study the impact of the vibrational mode lifetime $T_1$, as well as the detailed behaviour near the real scattering condition $(\omega_{\mathbf{k}}-\omega_{\mathbf{k}-\mathbf{q}})^2=\nu_{\mathbf{q}}^2$, in which the perturbative scheme of the BCS theory breaks down.

Figure~\ref{fig2}(b) compares qualitatively the numerical results around the real scattering condition with the perturbative BCS theory. Near the real scattering condition, the time scale $\Delta t = \hbar / \Delta E$ becomes comparable to $T_{\rm 1}$, and the correlation becomes non-monotonic, with a weak dependence on the exact value of $T_{\rm 1}$. In this neighborhood, the perturbative expansion breaks down and the BCS results are no longer reliable. At higher shifts the BCS branch deviates towards lower correlations due to the stretching vibrational modes [see Fig.~\ref{fig2}(c)], not included in the numerical simulation.

Figure~\ref{fig2}(b,c) also show the BCS solution for a simplified model of two vibrational modes. The sharp dip in $g^{(2)}_{S,aS}(0)$ reveals that the real processes in Fig.~\ref{fig1}(b) jeopardize the correlation. For several Raman modes, these dips superimpose leading to the complexity observed in Fig.~\ref{fig2}(a). This loss of correlation is not due to a weakening of the virtual process -- on the contrary, the gap in Eq.~(\ref{gap}) becomes largest at the vicinity of the real scattering condition. The validity of the BCS scheme suggested by the data in Fig.~\ref{fig2}(a) is therefore supported by the numerical results of Fig.~\ref{fig2}(b). Notice that even in the regime where real processes are present, virtual processes from other vibrational modes may cooperate, leading to important contributions to the total correlation.

\begin{figure}
\includegraphics[trim={0 5cm 0 3cm},clip,width=\columnwidth]{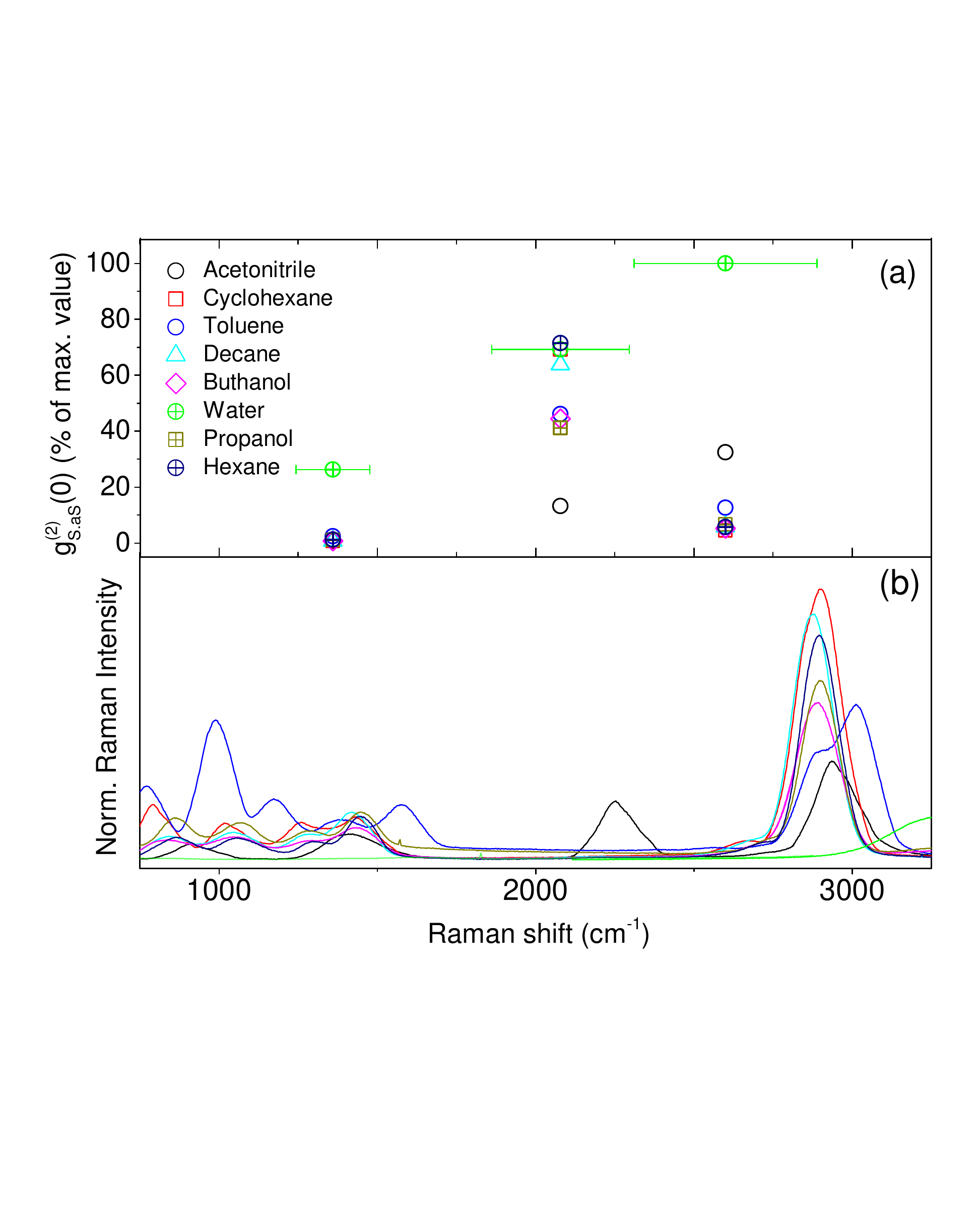}
\caption{Stokes-anti-Stokes correlation function $g^{(2)}_{S,aS}(0)$ for different media. (a) Experimental values for the different media are shown as different symbols (see legend) for three different Raman shifts. The percentage $g^{(2)}_{S,aS}(0)$ with respected to the highest observed value is shown, also corrected to account for the dependence of the measured $g^{(2)}_{S,aS}(0)$ with the band-overlap between Stokes and anti-Stokes filters. The horizontal bars indicate the overlap between bandpass filters, and are shown for only one set of data. (b) Stokes Raman spectra for the different media, following the same color pattern as in (a), all normalized to the highest intensity Raman peak. The unusually broad Raman peaks reflect the use of a femtosecond laser to excite the samples.}
\label{fig3}
\end{figure}

Our conclusion should hold for other transparent media, crystal or fluid. In fluids, clusters of molecules will vibrate simultaneously since a photon in a laser pulse cannot resolve a single molecule -- its wavelength is much longer than the molecular separation. Similarly, the coherence length of optical phonons in solids is typically much smaller than the light wavelength~\cite{beams}, so that the atoms oscillating in phase occupy a region much larger than the typical size of the phonons. Therefore this phenomenon is collective in either case. As shown in Fig.~\ref{fig3}, the results hold for acetonitrile, buthanol, cyclohexane, decane, hexane propanol and toluene. Notice that acetonitrile has the lowest $g^{(2)}_{S,aS}(0)$ value for the 2,100\,cm$^{-1}$ band-pass filter because it is the only sample with a Raman peak within this energy range. For the filter at 2,600\,cm$^{-1}$, water, acetonitrile and toluene have the highest $g^{(2)}_{S,aS}(0)$ values (in this order), consistent with their Raman peaks being shifted to higher frequencies, out of the filter band-pass range [see Fig.~\ref{fig3}(b)]. The generation of the Stokes-anti-Stokes pairs occurs via the Cooper pair mechanism, a quantum correlated phenomenon. The decrease in $g^{(2)}_{S,aS}(0)$ is related mostly to an increase in the background caused by accidental coincidences, rather then by the increase of the SaS-generation efficiency~\cite{SM}. An increase in the number of photons reaching the detector generates more accidental coincidences -- which illustrates the correspondence principle in this phenomenon, showing that at the limit of large quantum numbers the quantum correlations disappear and the dynamics becomes classical.

In conclusion, we demonstrated the formation of photonic Cooper-like pairs in transparent media in general via the virtual SaS process. At this point, it is not clear what materials properties influence the ultimate SaS production efficiency. This is a matter for future studies. Perhaps the most pressing question left open by these results is to what extent may the analogy between this phenomenon and superconductivity be drawn.  Remarkably similar phenomenology was already observed in non-linear media, such as the  dispersion cancellation~\cite{steinberg} and the electromagnetically induced transparency~\cite{imamoglu}. These could be interpreted as the photonic counterparts of supercurrents. While one may speculate about applications enabled by potential further analogies, the phenomenon at hand is already expected to be sufficient for the creation of pairs of entangled photons that can be energy tunable, or to detect vibrational modes that may be silent under regular Raman spectroscopy (due to selection rules) but measurable with SaS processes.

A.J. and F.S.A.J acknowledges financial support from CNPq (552124/2011-7, 460045/2014-8) and FINEP(01.13.0330.00). M.F.S. acknowledges CNPq (305384/2015-5).  B.K. acknowledges CNPq (304869/2014-7) and FAPERJ E-26/202.915/2015. A.S. acknowledges CNPq (309861/2015-2). R.M.S acknowledges FAPERJ (E-05/2016tTXE-05/2016). Correspondence and requests for materials should be addressed to A.S. (also@if.ufrj.br) and A.J. (adojorio@fisica.ufmg.br).


\begin{thebibliography}{99}
\bibitem{4wave} This process is completely distinct from four wave mixing (FWM). Some differences include the fact that FWM generates arbitrary pairs of photons with frequencies $f_3$ and $f_4$ as long as their sum equals the sum of incoming frequencies $f_1+f_2=f_3+f_4$. The virtual SaS processes only scatter pairs of photons from an input frequency $f_{in}$ into symmetric pairs $f_{in}\pm\delta$. Moreover, the exchanged phonons could carry momentum, adding a further variable to the non-linear effect. Finally, common FWM processes are directly generated by virtual (or real) electronic excitations. Virtual SaS processes include the generation of virtual electronic excitations, but the main ingredient for the interaction is the phonon, which is absent in FWM.
\end{thebibliography}
\end{document}